\documentclass[aps,prd,superscriptaddress,showpacs,showkeys]
{revtex4}
\def\bear{\begin{eqnarray}}
\def\ear{\end{eqnarray}}
\newcommand{\tr}{\text{tr}}
\newcommand{\Tr}{\text{Tr}}
\newcommand{\nF}{n_{\text{F}}}
\newcommand{\E}{\text{e}}
\begin{document}

\preprint{}
\title{Zero modes, beta functions and IR/UV interplay in higher-loop QED}
\author{Gerald V. Dunne}
\affiliation{Department of Physics, University of Connecticut,
Storrs, CT 06269-3046, USA}
\author{Holger Gies}
\affiliation{Theory Division, CERN, CH-1211 Geneva 23, Switzerland}
\author{Christian Schubert}
\affiliation{Instituto de F\'{\i}sica y Matem\'aticas,
Universidad Michoacana de San Nicol\'as de Hidalgo,
Apdo. Postal 2-82,
C.P. 58040, Morelia, Michoac\'an, M\'exico}
\affiliation{California Institute for Physics and Astrophysics,
366 Cambridge Ave., Palo Alto, CA 94306, USA}

\begin{abstract}
We analyze the relation between the short-distance behavior of quantum
field theory and the strong-field limit of the background field
formalism, for QED effective Lagrangians in self-dual backgrounds, at
both one and two loop. The self-duality of the background leads to zero
modes in the case of spinor QED, and these zero modes must be taken into
account before comparing the perturbative $\beta$ function coefficients
and the coefficients of the strong-field limit of the effective
Lagrangian. At one-loop this is familiar from instanton physics, but
we find that at two-loop the role of the zero modes, and the interplay
between IR and UV effects in the renormalization, is quite different. Our
analysis is motivated in part by the remarkable simplicity of the
two-loop QED effective Lagrangians for a self-dual constant background,
and we also present here a new independent derivation of these two-loop
results. 
\end{abstract}
\maketitle

\vspace{-10.4cm}

{\hfill CERN-TH/2002-252}
 
{\hfill UMSNH-Phys/02-10}

\vspace{+10cm}

\section{Introduction}
\label{introduction}
\renewcommand{\theequation}{1.\arabic{equation}}
\setcounter{equation}{0}

In quantum field theory there is a close connection between the
short-distance behavior of renormalized Green's functions and the
strong-field limit of associated quantities calculated using the
background field method. This phenomenon can be interpreted as 
an IR/UV connection in the sense that the ultraviolet (UV) and infrared
(IR) divergences are correlated. This correspondence
leads, for example, to a direct relation between the perturbative
$\beta$ function and the strong-field asymptotics of the effective
Lagrangian. Ritus derived this relation in the context of QED using
the renormalization group, with the assumption that the strong-field
limit of the renormalized effective Lagrangian is mass-independent
\cite{ritus1,ritus2,ritus3}. Another, equivalent, derivation which
invokes the scale anomaly \cite{crewther,anomaly,nielsen} in a massless
limit, has been given in various forms by many authors
\cite{matinian,pagels,leutwyler,minkowski,elizalde,fujikawa,hansson,giesbook}.
Also, other approaches have been developed for extracting the $\beta$
function from the effective Lagrangian, using either the operator product
expansion \cite{shifman} or the worldline formalism \cite{ss}.
These issues have most often been investigated for magnetic or
chromo-magnetic background fields, or for self-dual backgrounds (such
as instantons) at one loop. Here, in this paper, we re-examine these
issues at the two loop level for  self-dual background fields.
A self-dual background is special because it gives rise to zero modes
in a spinor theory \cite{schwarz}. Also, for a self-dual background it
is not possible to distinguish between the bare Lagrangian,
$F_{\mu\nu}F^{\mu\nu}$, and the other Lorentz invariant combination,
$F_{\mu\nu}\tilde{F}^{\mu\nu}$, which characterizes the zero-mode
contributions, so it is necessary to identify and separate the zero mode
contributions before taking the strong-field limit. Our analysis
concentrates on QED, and is motivated in part by the recent results that
the two-loop effective Lagrangian, in both spinor and scalar QED, for a
constant self-dual background field has a remarkably simple closed
form \cite{ds1,ds2,ds3}. However, since many of the
simplifications we find are due to the self-duality of the background,
rather than due to the precise form of the background, another
motivation is to learn which features might be applied to
higher-loop calculations in other self-dual systems, such as for
example QCD with instanton backgrounds.

In Section II we review the IR/UV correspondence between the
perturbative $\beta$ function and the strong-field limit of the
effective Lagrangian. We show that a naive application of this
correspondence fails for spinor QED with a self-dual background, at
both one loop and two loop. In Section III we show how this apparent
discrepancy is resolved at one loop by the separation of the zero
mode contribution to the effective Lagrangian. In Section IV we show
that at two loop the mechanism whereby the discrepancy is resolved is
rather different, coming instead from a zero mode contribution to the
mass renormalization. In this Section we also provide an
independent derivation of the two loop effective Lagrangians for a
constant self-dual background which were found previously
\cite{ds1,ds2,ds3} using the worldline formalism. The final section
contains our conclusions and an appendix describes the
calculation of the finite part of the mass renormalization.

\section{Strong-field limits and beta functions}
\label{strong}
\renewcommand{\theequation}{2.\arabic{equation}}
\setcounter{equation}{0}

\subsection{General Argument}
\label{general}

We begin by recalling the general argument
\cite{ritus1,ritus2,ritus3,matinian,pagels,leutwyler,minkowski,elizalde,fujikawa,hansson,giesbook}
relating the strong-field asymptotic behavior of the effective
Lagrangian to the perturbative $\beta$ function. 
As we are interested here in QED, we present the argument for an
abelian gauge theory, but it is more  general.
Consider an abelian gauge field coupled to spinor or
scalar matter fields, which are either explicitly massless or which have
a well-defined massless limit. Then the trace anomaly for the
energy-momentum tensor states that \cite{crewther,anomaly,nielsen}
\begin{eqnarray}
\langle \Theta^{\mu}_{\mu}\rangle = \frac{\beta(\bar{e})}{2\bar{e}}\,
\frac{e^2}{\bar{e}^2}\, (F_{\mu\nu})^2,
\label{anomaly}
\end{eqnarray}
where $\bar{e}$ is the running coupling, and $\beta(\bar{e})$ is the
$\beta$ function, defined below in (\ref{betadef}). The
expectation value of the energy-momentum tensor can also be related
to the effective Lagrangian for a constant background field strength
$F_{\mu\nu}$:
\begin{eqnarray}
\langle \Theta^{\mu\nu}\rangle =-\eta^{\mu\nu}\,{\cal L}_{\rm
eff}+2\frac{\partial {\cal L}_{\rm eff}}{\partial \eta_{\mu\nu}}.
\label{emom}
\end{eqnarray}
These two relations, (\ref{anomaly}) and (\ref{emom}), determine the
effective Lagrangian to be of the form
\begin{eqnarray}
{\cal L}_{\rm eff}=-\frac{1}{4}\, \frac{e^2}{\bar{e}^2(t)}\,
F_{\mu\nu}F^{\mu\nu},
\label{efflag}
\end{eqnarray}
where the "renormalization group time", $t$, is expressed in terms of the
scale set by the field strength, serving as the renormalization scale
parameter $\mu^2\sim e|F|$,
\begin{eqnarray}
t=\frac{1}{4}\,\ln\left(\frac{e^2 |F^2|}{\mu_0^4}\right),
\label{tscale}
\end{eqnarray}
and $\mu_0$ denotes a fixed reference scale at which, for example, the
value of the coupling may be measured.  Note that in this
argument the field strength plays the role which is usually
associated with a momentum transfer $Q^2$. 
This already suggests at a very
basic level why the strong-field and short-distance limits are
related.

The $\beta$ function is defined in terms of the running of the
coupling as 
\begin{eqnarray}
\beta(\bar{e}(t))&=&\frac{d\bar{e}(t)}{dt}.
\label{betadef}
\end{eqnarray}
To see how this solution (\ref{efflag}) leads to an explicit
connection between the strong-field asymptotics of ${\cal L}_{\rm
eff}$ and the perturbative
$\beta$ function, note that  (\ref{betadef}) can also be expressed as
\begin{eqnarray}
t=\int^{\bar{e}(t)}_e \, \frac{de^\prime}{\beta(e^\prime)},
\label{invdef}
\end{eqnarray}
where $e\equiv \bar{e}(0)$. Making a perturbative expansion of the
$\beta$ function
\begin{eqnarray}
\beta(e)=\beta_1 e^3 +\beta_2 e^5 +\dots
\label{pertbeta}
\end{eqnarray}
the relation (\ref{invdef}) determines the running coupling,
$\bar{e}(t)$, in terms of $e$ as
\begin{eqnarray}
\frac{1}{\bar{e}^2(t)}=\frac{1}{e^2}-2\beta_1\, t-2 \beta_2 e^2\, t+
O(e^4 t^2).
\label{ebar}
\end{eqnarray}
Inserting this into (\ref{efflag}), the strong-field asymptotics of the
effective Lagrangian is, to two-loop order, 
\begin{eqnarray}
{\cal L}_{\rm eff}\sim \frac{1}{16}\left(2\beta_1e^2+2 \beta_2
e^4+\dots\right) \, F_{\mu\nu}F^{\mu\nu}\,
\ln\left(\frac{e^2|F^2|}{\mu_0^4}\right),
\label{corr}
\end{eqnarray}
where, as is conventional, we have subtracted the classical Lagrangian,
$-\frac{1}{4}F_{\mu\nu}F^{\mu\nu}$, from ${\cal L}_{\rm eff}$.

In order to illustrate this correspondence explicitly, we recall that
the QED
$\beta$ functions, for spinor and scalar QED, to two-loop order, are
\begin{eqnarray}
\beta_{\rm spinor}
  &=&\frac{e^3}{12\pi^2}+\frac{e^5}{64\pi^4}+\dots
  \label{spbeta} \\
\beta_{\rm scalar}
  &=&\frac{e^3}{48\pi^2}+\frac{e^5}{64\pi^4}+\dots
  \label{scbeta}
\end{eqnarray}
The $\beta$ function can also be expressed in terms of
$\alpha=e^2/(4\pi)$ instead of $e$ by a change of variables,
$\beta^{(\alpha)}=\frac{d\alpha}{dt}=\frac{2e}{4\pi}
\beta(e)\Big|_{e=\sqrt{4\pi\alpha}}$, leading to the form
\begin{eqnarray}
\beta_{\rm spinor}^{(\alpha)}&=&
  \frac{2\alpha^2}{3\pi}+\frac{\alpha^3}{2\pi^2}+\dots
  \label{spbetaa}\\
\beta_{\rm scalar}^{(\alpha)}&=&
  \frac{\alpha^2}{6\pi}+\frac{\alpha^3}{2\pi^2}+\dots 
  \label{scbetaa}
\end{eqnarray}

\subsection{Explicit example: constant magnetic field background}

Equation (\ref{corr}) gives a direct correspondence between the
perturbative $\beta$ function coefficients and the strong-field
behavior of the effective Lagrangian. We now compare this with some
explicit results where the effective Lagrangian is known. First, consider
the Euler-Heisenberg effective Lagrangian for a constant background
{\it magnetic} field, of strength
$B$. At one loop, the on-shell renormalized effective Lagrangians, for
spinor and 
scalar QED, are \cite{eh,schwinger}
\bear
{\cal L}_{\rm spinor}^{(1)\,{\rm magnetic}}
&=&
-{e^2 B^2\over 8\pi^2}
\int_0^{\infty}{ds\over s}
\,e^{-m^2s/(eB)}
\biggl\lbrack
{1\over s\,\tanh(s)}-\frac{1}{s^2}-{1\over 3}
\biggr\rbrack,
\label{L1spinormag}
\ear
\bear
{\cal L}_{\rm scalar}^{(1)\,{\rm magnetic}}
&=&
{e^2 B^2\over 16\pi^2}
\int_0^{\infty}{ds\over s}
\,e^{-m^2s/(eB)}
\biggl\lbrack
{1\over s\,\sinh(s)}-\frac{1}{s^2}+{1\over 6}
\biggr\rbrack.
\label{L1scalarmag}
\ear
The leading strong-field asymptotics is determined by the IR
behavior of the propertime integrand for $s\to \infty$; at the
one-loop level this yields
\begin{eqnarray}
{\cal L}_{\rm spinor}^{(1)\,{\rm magnetic}}\sim 
\frac{e^2B^2}{24\pi^2}\ln\left(\frac{eB}{m^2}\right)+\dots
\label{1loopspinorasym}
\end{eqnarray}
\begin{eqnarray}
{\cal L}_{\rm scalar}^{(1)\,{\rm magnetic}}\sim 
\frac{e^2B^2}{96\pi^2}\ln\left(\frac{eB}{m^2}\right)+\dots
\label{1loopscalarasym}
\end{eqnarray}
Noting that $-\frac{1}{4}F_{\mu\nu}F^{\mu\nu}=-\frac{1}{2}B^2$, and
comparing with the correspondence (\ref{corr}), we deduce that
$\beta_1^{\rm spinor}=\frac{1}{12\pi^2}$ and $\beta_1^{\rm
scalar}=\frac{1}{48\pi^2}$, in agreement with
the one-loop
$\beta$ function coefficients quoted in (\ref{spbeta}) and (\ref{scbeta}).

The two-loop renormalized effective Lagrangians for
a constant background field were derived by Ritus for both spinor
\cite{ritus1} and scalar \cite{ritus2} QED. While the actual expressions
for the effective Lagrangians are complicated double parameter integrals,
it is nevertheless possible to extract the two-loop leading strong-field
asymptotics for the constant magnetic field case \cite{ritus3}:
\begin{eqnarray}
{\cal L}_{\rm spinor}^{(2)\,{\rm magnetic}}\sim 
\frac{e^4 B^2}{128\pi^4}\ln\left(\frac{eB}{m^2}\right)+\dots
\label{2loopspinorasym}
\end{eqnarray}
\begin{eqnarray}
{\cal L}_{\rm scalar}^{(2)\,{\rm magnetic}}\sim 
\frac{e^4 B^2}{128\pi^4}\ln\left(\frac{eB}{m^2}\right)+\dots
\label{2loopscalarasym}
\end{eqnarray}
Once again, comparing with the correspondence (\ref{corr}), 
we deduce that $\beta_2^{\rm spinor}=\beta_2^{\rm
scalar}=\frac{1}{64\pi^4}$, in agreement with 
the two-loop $\beta$ function coefficients quoted in (\ref{spbeta})
and (\ref{scbeta}). 

\subsection{Explicit example: constant self-dual background}

Another interesting solvable case is when the constant background field
is self-dual:
\bear
F_{\mu\nu}=\tilde
F_{\mu\nu}\equiv\frac{1}{2}\varepsilon_{\mu\nu\alpha\beta}
F^{\alpha\beta}.
\ear
It is well-known that self-dual backgrounds have special properties
which often lead to dramatic simplifications. This can be traced to
the fact that a self-dual background has definite helicity and the
Dirac operator in such a background has a quantum mechanical
supersymmetry \cite{thooft,dadda,duff}. Since a self-dual
background has definite helicity, the effective Lagrangian for such
a background can be used as a generating functional for amplitudes
with all external lines having the same helicity. It is also well-known
that many remarkable simplifications occur for such helicity
amplitudes \cite{mahlon,bardeen}. Recently it has been found that
analogous simplifications occur in the two-loop effective Lagrangian
itself
\cite{ds1,ds2,ds3}.

At one-loop, the on-shell renormalized effective Lagrangians for a
constant self-dual background can be deduced from the results of Euler and
Heisenberg
\cite{eh} and Schwinger
\cite{schwinger}:
\bear
{\cal L}_{\rm spinor}^{(1)\,{\rm self-dual}}
=
-{e^2f^2\over 8\pi^2}
\int_0^{\infty}
{ds\over s}\,e^{-m^2 s/(ef)}
\biggl[
{\coth^2s}-\frac{1}{s^2}-\frac{2}{3}
\biggr],
\label{1lspinorsd}
\ear
\bear
{\cal L}_{\rm scalar}^{(1)\,{\rm self-dual}}=
\frac{e^2f^2}{16\pi^2}\int_0^\infty
\frac{ds}{s}\, e^{-m^2
s/(ef)}\,\left[\frac{1}{\sinh^2 s}-\frac{1}{s^2}+\frac{1}{3}
\right].
\label{1lscalarsd}
\ear
Here, $f$ denotes the magnitude of the self-dual field strength,
\begin{equation}
\frac{1}{4}F_{\mu\nu}F^{\mu\nu}\equiv f^2.
\end{equation} 
Note that the one-loop renormalized
effective Lagrangians in (\ref{1lspinorsd}) and (\ref{1lscalarsd}) satisfy
${\cal L}^{(1)\,{\rm self-dual}}_{\rm spinor}=-2{\cal L}^{(1)\,{\rm
self-dual}}_{\rm scalar}$, which is a consequence of the supersymmetry of
the self-dual background
\cite{dadda,duff}.

At two-loop, the on-shell renormalized effective Lagrangians for a
constant self-dual background can be expressed in closed-form
\cite{ds1,ds2,ds3} in terms of the digamma function,
\begin{eqnarray}
{\cal L}_{\rm spinor}^{(2)\,{\rm self-dual}}
=
-2\alpha \,{m^4\over (4\pi)^3}\frac{1}{\kappa^2}\left[
3\xi^2 (\kappa)
-\xi'(\kappa)\right],
\label{2lsp}
\end{eqnarray}
\begin{eqnarray}
{\cal L}_{\rm scalar}^{(2)\,{\rm self-dual}}
=
\alpha \,{m^4\over (4\pi)^3}\frac{1}{\kappa^2}\left[
{3\over 2}\xi^2 (\kappa)
-\xi'(\kappa)\right].
\label{2lsc}
\end{eqnarray}
Here the dimensionless parameter $\kappa$ is defined as
\begin{eqnarray}
\kappa=\frac{m^2}{2ef},
\label{kappa}
\end{eqnarray}
and the function $\xi(\kappa)$ is
\begin{eqnarray}
\xi(\kappa)=-\kappa\left(\psi(\kappa)-\ln \kappa+\frac{1}{2\kappa}\right).
\label{xi}
\end{eqnarray}
Note that this function $\xi(\kappa)$ is essentially the digamma function,
$\psi(\kappa)=\frac{d}{d\kappa}\ln\Gamma(\kappa)$, with the first two
terms of its large $\kappa$ asymptotic expansion subtracted (see Eq
(6.3.18) in \cite{abramowitz}). It is interesting to note also that the
one-loop expressions (\ref{1lspinorsd}) and (\ref{1lscalarsd}) can be
expressed \cite{ds1} simply in terms of the function $\int^\kappa \xi$.

From (\ref{kappa}) it is clear that the strong-field limit
corresponds to the small $\kappa$ limit. Thus, the strong-field behaviors
can be deduced from the known series expansion of $\psi(\kappa)$ (see Eq
(6.3.14) in \cite{abramowitz}).
For scalar QED in a constant self-dual background, the one- and two-loop
leading strong-field behaviors of the effective Lagrangians
(\ref{1lscalarsd}) and (\ref{2lsc}) are
\cite{ds1}
\begin{eqnarray}
{\cal L}^{(1)\,{\rm self-dual}}_{\rm scalar}\sim
\frac{e^2}{48\pi^2}\,f^2\,\ln\left(\frac{ef}{m^2}\right)+\dots
\label{1lscalarsdasym}
\end{eqnarray}
\begin{eqnarray}
{\cal L}^{(2)\,{\rm self-dual}}_{\rm scalar}\sim
\frac{e^4}{64\pi^4}\,f^2\,\ln\left(\frac{ef}{m^2}\right)+\dots
\label{2lscalarsdasym}
\end{eqnarray}
Once again, comparing with (\ref{corr}), we see that
the coefficients of this leading behavior agree with the
scalar QED $\beta$ function coefficients at one- and two-loop in
(\ref{scbeta}), as was noted already in \cite{ds1}.

On the other hand, for spinor QED in a constant self-dual background, the
leading strong-field behaviors of the effective Lagrangians
(\ref{1lspinorsd}) and (\ref{2lsp}) are 
\begin{eqnarray}
{\cal L}^{(1)\,{\rm self-dual}}_{\rm spinor}\sim
- \frac{e^2}{24\pi^2}\,f^2\,\ln\left(\frac{ef}{m^2}\right)+\dots
\label{1lspinorsdasym}
\end{eqnarray}
\begin{eqnarray}
{\cal L}^{(2)\,{\rm self-dual}}_{\rm spinor}\sim -
\frac{e^4}{32\pi^4}\,f^2\,\ln\left(\frac{ef}{m^2}\right)+\dots
\label{2lspinorsdasym}
\end{eqnarray}
Comparing with (\ref{corr}), we see that
the coefficients of these leading behaviors {\bf do not} agree with the
spinor QED $\beta$ function coefficients in (\ref{spbeta}), at either
one-loop or two-loop. This apparent mis-match is the issue which will
be resolved in the following sections of this paper. As already hinted
in the Introduction, the key is that for the spinor case (but not the
scalar case) there are zero modes in a self-dual background, and
these zero modes must be separated first, before making the
strong-field comparison.

\section{One-loop Euler-Heisenberg Lagrangian for a 
self-dual field}
\label{oneloopselfdual}
\renewcommand{\theequation}{3.\arabic{equation}}
\setcounter{equation}{0}

The results of (\ref{1lspinorsdasym}) and
(\ref{2lspinorsdasym}) make it clear that the general argument given
in Section \ref{general} needs to be modified in some way for the case
of spinor QED with a self-dual background. This modification must
take account of the presence of zero modes for spinor QED in a
self-dual background. Furthermore, this must be done at any loop
order. At one-loop it is well-known that the small mass limit is
complicated by the existence of normalizable zero modes for the
massless Dirac equation, and that the resolution is known to involve
the separation of a logarithmic term proportional to the number of
zero modes \cite{thooft,schwarz,ck,carlitz}. Here we are 
interested in the role of the zero modes at two loops.
However, in order to proceed to the two-loop level in subsequent
sections, we first briefly recap the evaluation and renormalization of
the one-loop Euler-Heisenberg effective Lagrangian for a self-dual
background. This will serve to explain the difference between scalar
and spinor QED, in a self-dual background, 
with respect to the connection between the
$\beta$ function coefficients and the strong-field behavior of the
on-shell renormalized effective Lagrangian.

\subsection{One-loop scalar QED in a self-dual background}

For scalar QED, the one-loop effective Lagrangian, ${\cal L}^{(1)}$,
is defined as
\bear
\int d^4x\,{\cal L}_{\rm scalar}^{(1)}=-\frac{1}{2}\,
\ln\det\left(-D^2+m^2\right),
\label{1lsclag}
\ear
where $D_\mu=\partial_\mu+ie A_\mu$ is the covariant derivative.
This can also be expressed in terms of the scalar propagator,
$G=\frac{1}{-D^2+m^2}$, for scalar particles in the given  
background. For a self-dual background, the scalar
propagator has a simple position space representation (up to an
unimportant gauge dependent phase),
\bear
G(x,x^\prime)=\left(\frac{ef}{4\pi}\right)^2\int_0^\infty
\frac{dt}{\sinh^2(eft)}\, \exp\left[-m^2 t-\frac{ef}{4}(x-x^\prime)^2
\coth(eft)\right].
\label{scprop}
\ear
The unrenormalized effective Lagrangian is therefore
\bear
{\cal L}_{\rm scalar,unren.}^{(1)\,{\rm
self-dual}}=\left(\frac{ef}{4\pi}\right)^2 \int_0^\infty
\frac{dt}{t}\, e^{-m^2
t}\,\left[\frac{1}{\sinh^2(eft)}
\right].
\label{1lscunren}
\ear 
To obtain the renormalized one-loop effective Lagrangian, ${\cal
L}^{(1)}_R$, we first subtract the (divergent) zero-field
contribution, and then introduce  an ultraviolet cutoff $\Lambda$
through a lower bound $\frac{1}{\Lambda^2}$ on the proper-time $t$
integral. Furthermore, we introduce a (redundant) renormalization
scale $\mu$ by writing
\begin{eqnarray}
{\cal L}_{\rm scalar}^{(1)\,{\rm self-dual}}
&=&\left(\frac{ef}{4\pi}\right)^2\int_0^\infty
\frac{dt}{t}\, e^{-m^2
t}\,\left[\frac{1}{\sinh^2(eft)}-\frac{1}{(eft)^2}+\frac{1}{3}
  \right]+
\frac{e^2}{48\pi^2}\,f^2\left[ \ln\left(\frac{m^2}{\mu^2}\right)+\gamma+
\ln\left(\frac{\mu^2}{\Lambda^2}\right)\right]
\nonumber\\
&=&{\cal L}_{\rm
    scalar, R}^{(1)\,{\rm self-dual}} +\frac{\alpha}{12\pi}\,
f^2\left[\ln\left(\frac{\mu^2}{\Lambda^2}\right) +\gamma\right].
\label{1lscren}
\end{eqnarray}
Here $\gamma$ is Euler's constant, and we dropped terms of ${\cal O}(m^2/\Lambda^2)$. In the last line,
we separated the renormalized one-loop Lagrangian, ${\cal
L}^{(1)}_R$, from the counterterm. The latter can be combined with
the unrenormalized classical action, corresponding to charge and field
strength renormalization, so that
$\alpha=\alpha(\mu)=\frac{e^2(\mu)}{4\pi}$ becomes the running coupling
\cite{schwinger}. For instance, implementing electron mass-shell
renormalization conditions, $\mu=m$, so that
$\alpha(m)\simeq1/137$, the $\ln m^2/\mu^2$ term would drop out.
However, since we are interested in the strong-field limit and its
mass-(in-)dependence, let us keep the $\mu$ dependence. 

For our purposes, it is important to observe that  the strong-field
limit of the effective Lagrangian [see Eq.
(\ref{1lscalarsdasym})] comes from the
$\frac{1}{3}$ term inside the square brackets in the first line of
(\ref{1lscren}), while the one-loop
$\beta$ function coefficient is determined by the $\mu$ dependence in
the logarithmic term on the second line of (\ref{1lscren}), which is
the charge renormalization counterterm. These terms have the same
coefficient, which illustrates the connection (\ref{corr}) between the
strong-field limit of the one-loop effective Lagrangian and the
one-loop $\beta$ function. It also confirms the assumption
\cite{ritus3} of mass-independence of the strong-field limit, since
in the $m\to 0$ limit we
can write
\begin{equation}
{\cal L}_{\rm scalar, R}^{(1)\,{\rm self-dual}}
\sim \frac{\alpha}{12\pi}\,f^2\ln\left(\frac{ef}{m^2}\right) 
+\frac{\alpha}{12\pi}\,f^2\ln\left(\frac{m^2}{\mu^2}\right) 
=\frac{\alpha}{12\pi}\,f^2\ln\left(\frac{ef}{\mu^2}\right), 
\label{1lscmi}
\end{equation}
which guarantees that the limit $m\to 0$ can be taken. This is
important because the existence of a well-defined massless limit is a
necessary prerequisite for the trace-anomaly argument described in
Section IIA.

\subsection{One-loop spinor QED in a self-dual background}

For spinor QED, the one-loop effective Lagrangian, ${\cal L}^{(1)}$,
is defined as
\bear 
\int d^4x\, {\cal  L}_{\rm spinor}^{(1)}=\ln\det\left(D\hskip-7pt
  /\hskip 5pt 
  +m\right)=\frac{1}{2}\ln\det\left(-D\hskip -7pt /\hskip
  5pt^2+m^2\right).
\label{1lsplag}
\ear
A self-dual background has definite helicity
\cite{thooft,schwarz,dadda,duff}, which has the consequence
that
\bear
D\hskip -7pt /\hskip 5pt^2\, P_L=D^2\, P_L,
\label{magic}
\ear
where $P_L=\frac{1}{2}(1+\gamma_5)$ is the projector onto positive
helicity states. It follows that the spinor propagator $S$ can be
expressed in terms of the scalar propagator $G$
\cite{ck,carlitz,leeleepac},
\bear
S&=& \frac{1}{D\hskip -7pt /\hskip 5pt+m}\nonumber\\
&=&-\left(D\hskip -7pt
/\hskip 5pt-m\right) G\,P_L-G\, D\hskip -7pt /\hskip 5pt
P_R+\frac{1}{m}\left(1+D\hskip -7pt /\hskip 5pt G D\hskip -7pt /\hskip
5pt\right)\, P_R,
\label{spinorscalar}
\ear
where $P_R=\frac{1}{2}(1-\gamma_5)$.
This can also be expressed in proper-time form as \cite{leeleepac}
\bear
S=-\int_0^\infty dt\,e^{-m^2 t}\left[\left(D\hskip -7pt /\hskip
5pt-m\right) e^{D^2 t}\,P_L+e^{D^2 t}\, D\hskip -7pt /\hskip 5pt P_R-m
D\hskip -7pt /\hskip 5pt\frac{1}{D^2}e^{D^2 t} D\hskip -7pt /\hskip
5ptP_R-m \,P\right].
\label{ptspinorprop}
\ear
The last term in (\ref{ptspinorprop}) involves the projector, $P$, onto
the zero modes,
\bear
P=\left(1+D\hskip -7pt /\hskip 5pt G_0 D\hskip -7pt /\hskip
5pt\right)\, P_R,
\label{zmprojector}
\ear
where $G_0=\lim_{m\to0}G$ denotes the massless scalar propagator.
After some straightforward Dirac traces, one finds that the
one-loop spinor effective Lagrangian (\ref{1lsplag}) can be expressed
in terms of the scalar effective Lagrangian, up to a zero-mode
projection contribution,
\bear
{\cal L}^{(1)\,{\rm self-dual}}_{\rm spinor}=-\frac{1}{2} \frac{1}{V}
 \int_0^\infty
\frac{dt}{t}\, e^{-m^2 t}\left\{ \Tr_x\left[4\, e^{D^2 t}\right]+\Tr_x\,
\tr_{\rm Dirac}\, P\right\},
\label{separation}
\ear
where $V$ denotes the spacetime volume. The first term on the RHS
of Eq. (\ref{separation}) is just $-2 {\cal L}_{\rm
scalar}^{(1)\,{\rm self-dual}}$, which is, as mentioned above, a direct
reflection of the supersymmetry of the self-dual background at one-loop.
The second term on the RHS of Eq. (\ref{separation})  counts
the number (density) of zero modes 
\bear
\nF&\equiv&\frac{1}{V}\,\Tr_x\, \tr_{\rm Dirac} P\nonumber\\
&=&\left(\frac{ef}{2\pi}\right)^2.
\label{zmnumber}
\ear
Note that $\nF$ is just the square of the usual 2d Landau degeneracy
factor, since the 4d self-dual system factorizes into two orthogonal 2d
Landau systems \cite{leutwyler}.

Thus, the one-loop spinor effective Lagrangian can be expressed in
proper-time form as
\bear
{\cal L}^{(1)\,{\rm self-dual}}_{\rm
spinor}&=&-\frac{(ef)^2}{8\pi^2}\int_{1/\Lambda^2}^\infty
\frac{dt}{t} e^{-m^2
t}\left[\frac{1}{\sinh^2(eft)}-\frac{1}{t^2}+1-\frac{2}{3}
+\frac{2}{3}\right]\nonumber\\
&=&-\frac{(ef)^2}{8\pi^2}\int_0^\infty
\frac{dt}{t}\, e^{-m^2
t}\left[\frac{1}{\sinh^2(eft)}-\frac{1}{t^2}+\frac{1}{3}\right]
+\frac{e^2}{12\pi^2}f^2 \left[\ln\left(\frac{m^2}{\mu^2}\right)+\gamma+
 \ln\left(\frac{\mu^2}{\Lambda^2}\right)\right]
\nonumber\\
&=&{\cal L}^{(1)\,{\rm self-dual}}_{\rm spinor,R}+
\frac{\alpha}{3\pi}\,f^2 \left[\ln\left(\frac{\mu^2}{\Lambda^2}\right)
  +\gamma\right]. 
\label{1lspsdren}
\ear
On the first line of (\ref{1lspsdren}), the $1$ refers to the zero
mode contribution, and $\frac{2}{3}$ is added and subtracted to
achieve the charge renormalization. As in the scalar case
(\ref{1lscren}), the strong-field limit is read off from the last
term, $\frac{1}{3}$, inside the square brackets on the second line of
(\ref{1lspsdren}), while the one-loop $\beta$ function coefficient is
read off from the coefficient of the logarithmic terms, which  are
responsible for charge renormalization. In this spinor case, in
contrast to the scalar case, the coefficients of these two terms are
different. It is clear that the source of the difference is 
precisely the zero mode contribution. This explains the mis-match, at
one-loop, between the strong-field asymptotics of the one-loop
effective Lagrangian and the one-loop $\beta$ function coefficient in
the spinor case with the self-dual background field.  This mis-match,
due to the zero mode contribution, also prevents the strong-field
limit from becoming mass-independent [as it was in the scalar case:
see Eq. (\ref{1lscmi})], and therefore violates the assumptions of
the trace-anomaly argument given in Section IIA. 
  
There is another useful perspective on this mis-match in the one-loop
spinor self-dual case: note that the {\em unrenormalized} one-loop
Lagrangian in the scalar case (\ref{1lscunren}) is infrared (IR)
finite even in the massless limit. However, the {\em renormalized}
one-loop Lagrangian (\ref{1lscren})
has an IR divergence at the upper bound of the proper-time integral
  in the massless limit (which is actually cancelled by the $\ln
  m^2/\mu^2$ term). It is precisely the charge renormalization
  subtraction removing the logarithmic UV divergence which introduces
  the IR divergence. And it is the IR divergence which dominates the
  strong-field limit. Hence, we observe an UV/IR connection in QED:
  the strong-field limit results from the IR behavior of the
  proper-time integral which receives contributions from the
counter-terms controlling the UV behavior. This means that if the {\em
unrenormalized}
Lagrangian is IR finite (as it is in magnetic backgrounds for both
scalar and spinor QED, and in the scalar self-dual case), the $\beta$
function and the strong-field limit coincide. The presence of zero
modes, in the spinor self-dual case, obviously spoils this UV/IR
connection, because it leads to an additional IR divergence of the
unrenormalized Lagrangian.

\section{Two-loop Euler-Heisenberg Lagrangian for a self-dual field}
\label{twoloopselfdual}
\renewcommand{\theequation}{4.\arabic{equation}}
\setcounter{equation}{0}

We now turn to a two-loop analysis of the Euler-Heisenberg Lagrangian
for a self-dual field in both spinor and scalar QED. We 
concentrate again on the role of the zero modes. Interestingly, this role
will turn out to be of a different nature than at one loop. The
unrenormalized two-loop spinor Lagrangian can be written in
coordinate space as
\begin{equation}
{\cal  L}_{\rm spinor}^{(2)} =\frac{e^2}{2} \int d^4 x'\, {\cal
  D}(x-x')\, \tr_{\text{Dirac}}\, \bigl[ \gamma_\mu \langle
  x|S|x'\rangle \gamma_\mu \langle x'|S|x\rangle\bigr],
  \label{2lsplag}
\end{equation}
where we have introduced the photon propagator ${\cal
  D}(x-x')=[4\pi^2(x-x')^2]^{-1}$ and work in the Feynman gauge for
convenience. Furthermore, we have used bracket notation for the
propagators, $S(x,x')\equiv\langle x|S|x'\rangle$.
  
The influence of the zero-mode contribution can conveniently be
studied with the aid of representation
(\ref{spinorscalar},\ref{ptspinorprop}) of the spinor propagator $S$,
which allows for a separation of the zero-mode contribution,
\begin{equation}
{\cal  L}_{\rm spinor,\,z.m.}^{(2)} =2e^2 \int d^4 x'\, {\cal
  D}(x-x')\,\int_0^\infty dt\, \E^{-m^2 t}\, \langle x|\E^{D^2
  t}|x'\rangle\, \tr_{\text{Dirac}} \langle x'|P|x\rangle.
  \label{2lsplagzm}
\end{equation}
Here the first matrix element corresponds to the proper-time
integrand of the massless scalar propagator (\ref{scprop}), and the
second matrix element contains the projector $P$ onto the zero modes,
defined in (\ref{zmprojector}). For a constant self-dual background
\bear
\langle x |e^{D^2 t}|x^\prime \rangle
=\frac{1}{(4\pi)^2}\frac{(ef)^2}{\sinh^2(eft)}\,
\exp\left[-\frac{ef}{4}(x-x^\prime)^2 \coth(eft)\right].
\label{propexp}
\ear
Thus, we can write
\begin{equation}
{\cal  L}_{\rm spinor,\,z.m.}^{(2)} =\frac{e^2}{32\pi^4}\,(ef)
\int_0^\infty \frac{ds}{\sinh^2s} e^{-m^2 s/(ef)}
\int d^4 x'\frac{1}{(x-x^\prime)^2}\,
\exp\left[-\frac{ef}{4}(x-x^\prime)^2 \coth s\right]\,
\tr_{\text{Dirac}} \langle x'|P|x\rangle.
  \label{2lsplagzm1a}
\end{equation}
In the strong-field limit, $\frac{ef}{m^2}\to \infty$, we observe that
\begin{equation} 
\frac{1}{4\pi^2 (x-x')^2}\, \exp\left[-\frac{ef}{4}(x-x')^2 \coth
s\right]\,\to
\frac{1}{ef \coth s} \delta(x-x').
\label{strfid}
\end{equation}
Therefore, in the strong-field limit the zero mode contribution is
\begin{equation}
{\cal  L}_{\rm spinor,z.m.}^{(2)} \to \frac{\alpha}{2\pi} \int_0^\infty
ds\, \frac{\E^{-\frac{m^2}{ef}s}}{\sinh s\cosh s}\,\nF.
\label{2lsplagzm2}
\end{equation}
Our main observation here is that the zero-mode contribution
(\ref{2lsplagzm2}) to the unrenormalized two-loop Lagrangian is IR
finite even in the massless limit. 
This should be contrasted with the one-loop spinor case
[see Eqs. (\ref{separation}) and (\ref{zmnumber})], where the zero-mode
contribution is the source of the IR divergence. At two-loops, even
though
Eq.~(\ref{2lsplagzm2}) has a UV divergence to be absorbed in charge
renormalization, the corresponding subtraction contributes equally to the
$\beta$ function and the strong-field limit by virtue of the UV/IR
connection discussed at the end of Section III.
Therefore, the zero-mode
contribution identified in Eq.~(\ref{2lsplagzm2}) is {\bf not} the source
of the difference between the
$\beta$ function coefficients and the strong-field limit coefficients.

In order to pinpoint the actual source of the mis-match at the two-loop
level, let us perform the calculation in a straightforward way, starting
from Eq.~(\ref{2lsplag}) and using the relations
(\ref{spinorscalar}) and (\ref{ptspinorprop}) to trade the spinor
propagators $S$ for a representation in terms of the scalar
propagator $G$. (This derivation complements the two-loop derivations
in \cite{ds1,ds2}, which were done using the world-line formalism.) 
After taking the Dirac trace, we arrive at a simple form in terms of
matrix elements of $G$ \cite{ds2}:
\begin{equation}
{\cal  L}_{\rm spinor}^{(2)} =\frac{e^2}{2} \int d^4 x'\, {\cal
  D}(x-x')\, \Bigl[ -8\, \langle x|D_\alpha G|x'\rangle \langle
  x'|D_\alpha G|x\rangle +16\, \langle x|G|x'\rangle \langle
  x'|D_\alpha G D_\alpha|x\rangle+16\, \langle x|x'\rangle \langle
  x'|G|x\rangle \Bigr]. 
\label{2lspg}
\end{equation}
The last term corresponds to a ``tadpole'' diagram in the scalar
language, suggesting a quadratic divergence. However, this is only
seemingly the case; in fact, there is a cancellation of the last
term with a corresponding divergence in the second term. This can be seen
from the identity
\begin{equation}
\langle x|D_\alpha G D_\alpha|x'\rangle =-\langle x|x'\rangle+m^2
\langle x|G|x'\rangle +\frac{(ef)^2}{2} (x-x')^2 \langle
x|G|x'\rangle, \label{id}
\end{equation}
which makes it clear that the ``tadpole'' terms cancel in spinor
QED (as they should). 

The scalar two-loop Lagrangian is also written in a form analogous to
Eq.~(\ref{2lspg}), but with different coefficients:
\begin{equation}
{\cal  L}_{\rm scalar}^{(2)} =-e^2 \int d^4 x'\, {\cal
  D}(x-x')\, \Bigl[  \langle x|D_\alpha G|x'\rangle \langle
  x'|D_\alpha G|x\rangle + \langle x|G|x'\rangle \langle
  x'|D_\alpha G D_\alpha|x\rangle+4\, \langle x|x'\rangle \langle
  x'|G|x\rangle \Bigr]. \label{2lscg}
\end{equation}
In this case, the ``tadpole'' terms no longer cancel, and a
quadratic divergence remains. This is exactly as expected in the scalar
case, since this quadratic divergence reflects the presence of a relevant
operator, the scalar mass term, in scalar QED. In fact, as we shall
confirm below, the complete tadpole term, including its divergence, can be
absorbed into the mass renormalization.  

Before we proceed with the evaluation of these expressions, let us
comment that we derived Eqs.(\ref{2lspg},\ref{2lscg}) without recourse to
the explicit form of the background field. Only the self-duality of the
background has been used, so that we expect the similarities between
Eqs.(\ref{2lspg}) and (\ref{2lscg}) to hold also in a more general
context.

\subsection{Calculation of the two-loop Lagrangians}

The representations (\ref{2lspg}) and (\ref{2lscg}) of the spinor and 
scalar two-loop Lagrangians can be constructed from two basic terms (we
drop ``tadpole'' terms in the scalar case from now on, since they only
modify the mass renormalization as discussed above):
\begin{equation}
{\cal L}^{(2)}=\frac{\alpha^2}{(4\pi)^2}\, f^2\, \Bigl( A\, I_1 +B\,
I_2\Bigr), \label{decomp}
\end{equation}
where the numerical coefficients are $A=-4$ and $B=8$ in the spinor case,
and $A=B=-1$ in the scalar case. The integrals $I_1$ and $I_2$ are
\begin{eqnarray}
I_1&:=& \frac{(4\pi)^3}{\alpha f^2} 
  \int d^4x'\, {\cal D}(x-x')\, \langle x|D_\alpha G|
  x'\rangle \langle x'|D_\alpha G| x\rangle \nonumber\\
&=&\int_0^\infty y dy\, \E^{-\frac{m^2}{ef}y} \left[ \frac{2}{\sinh^2
    y} -\frac{\coth y}{\sinh^2 y} \int_0^1 du \bigl( \coth yu+ \coth
  y(1-u)\bigr) \right], \label{i11}\\
I_2&:=&  \frac{(4\pi)^3}{\alpha f^2} 
  \int d^4x'\, {\cal D}(x-x')\, \Bigl(\langle x| G|
  x'\rangle \langle x'|D_\alpha G D_\alpha| x\rangle +
  \langle x|G|x'\rangle \langle x|x '\rangle\Bigr)\nonumber\\
&=&\int_0^\infty y dy\, \E^{-\frac{m^2}{ef}y} 
  \left[ \frac{2}{\sinh^2 y} 
    +\frac{m^2}{ef}\frac{1}{\sinh y} \int_0^1 \frac{du}{ (\sinh yu)
(\sinh
       y(1-u))} \right], \label{i21}
\end{eqnarray}
where we have inserted the proper-time form of the scalar propagator
(\ref{scprop}), leading to a proper-time double integral.
Furthermore, we have rescaled the proper-time parameter as
$s=eft$, and then performed the substitutions
$y=s+s'$, $u=s'/(s+s')$. Both integrals $I_1$ and $I_2$ are IR finite but
UV divergent and require regularization. As at one loop, we introduce an
UV proper-time cutoff for each proper-time integral which implies
\begin{equation}
\int_0^\infty dy \to \int_{\frac{2ef}{\Lambda^2}}^\infty dy \quad ; \quad
\int_0^1 du \to \int_{\frac{ef}{\Lambda^2 y}}^{1-\frac{ef}{\Lambda^2
    y}} du.  
\end{equation}
Now the $u$ integrations can be performed, and we arrive at 
representations for $I_1$ and $I_2$ which are similar:
\begin{eqnarray}
I_1&=&\int_{\frac{2ef}{\Lambda^2}}^\infty dy\, \E^{-\frac{m^2}{ef} y} 
  \left\{\frac{2y}{\sinh^2 y} 
    + \frac{m^2}{ef} \frac{1}{\sinh^2 y} 
      \left(\ln\left[\frac{\sinh
(y-\frac{ef}{\Lambda^2})}{\sinh(\frac{ef}{\Lambda^2})}\right]-
\frac{ef}{m^2}\coth(y-\frac{ef}{\Lambda^2})\right) 
  \right\}, \label{i12}\\
I_2&=&\int_{\frac{2ef}{\Lambda^2}}^\infty dy\, \E^{-\frac{m^2}{ef} y} 
  \left\{\frac{2y}{\sinh^2 y} 
    + \frac{2m^2}{ef} \frac{1}{\sinh^2 y}\, 
      \ln\left[\frac{\sinh
(y-\frac{ef}{\Lambda^2})}{\sinh(\frac{ef}{\Lambda^2})}\right]
  \right\}. \label{i22}
\end{eqnarray}
An important observation is that each term in these
expressions for $I_1$ and $I_2$ can be naturally expressed in
terms of the function
$\xi(\kappa)$ which was defined in (\ref{kappa}) and (\ref{xi}). This
function $\xi(\kappa)$ has the following integral representation 
\begin{eqnarray}
\xi&=&-\frac{1}{2} \int_0^\infty dy\, \E^{-\frac{m^2}{ef}y}
  \left(\frac{1}{\sinh^2 y}-\frac{1}{y^2} \right)
\label{intrep} 
\end{eqnarray}
Recalling that $\kappa=m^2/(2 ef)$, it follows that
\begin{eqnarray}
\xi'&=&\int_0^\infty dy\, \E^{-\frac{m^2}{ef}y}\, y
  \left(\frac{1}{\sinh^2 y}-\frac{1}{y^2} \right)
\label{intrepprime}
\end{eqnarray}
Thus, the first term in each of $I_1$ and $I_2$ can be expressed in
terms of $\xi^\prime$:
\begin{eqnarray}
\int_{\frac{2ef}{\Lambda^2}}^\infty dy\, \E^{-\frac{m^2}{ef} y}\,
\frac{y}{\sinh^2 y}&=&\int_{\frac{2ef}{\Lambda^2}}^\infty dy\, \E^{-\frac{m^2}{ef} y}\,y\left(
\frac{1}{\sinh^2 y}-\frac{1}{y^2}\right)+
\int_{\frac{2ef}{\Lambda^2}}^\infty dy\, \E^{-\frac{m^2}{ef} y}\,
\frac{1}{y}\nonumber\\
&=&\xi^\prime(\kappa)+\left(-\gamma-\ln\left(\frac{2m^2}{\Lambda^2}
\right)\right)
\label{firstterm}
\end{eqnarray}
where we have dropped terms that vanish as the cutoff
is removed (i.e., as $\Lambda\to\infty$).

Similarly, by considering the integral representation for $\xi^2(\kappa)$
we find that the log terms in the expressions (\ref{i12}) and
(\ref{i22}) for $I_1$ and $I_2$ can also be expressed in terms of
$\xi(\kappa)$ as
\begin{eqnarray}
\frac{m^2}{ef}\int_{\frac{2ef}{\Lambda^2}}^\infty dy\,
\E^{-\frac{m^2}{ef} y} 
   \frac{1}{\sinh^2 y}\, 
      \ln\left[\frac{\sinh
(y-\frac{ef}{\Lambda^2})}{\sinh(\frac{ef}{\Lambda^2})}\right]
   &=&\frac{1}{2} -2\xi^2(\kappa)-2\frac{m^2}{ef} \left[ \ln
\left(\frac{\Lambda^2}{m^2}\right)+1-\gamma\right]\xi(\kappa)
\nonumber\\ &&-
2\kappa^2\left[\left(\ln\left(\frac{\Lambda^2}{m^2}\right)+1
-\gamma\right)^2+1-
\frac{2\Lambda^2}{m^2}\,\ln 2\right]
\label{idd}
\end{eqnarray}
where once again we have dropped terms which vanish as the cutoff
$\Lambda$ is removed. Also notice that the last parenthesis
term in (\ref{idd}) is proportional to $\kappa^2$,
and so when inserted into the two-loop effective Lagrangian
in (\ref{decomp}) this term gives a field-independent contribution to the
effective Lagrangian. Thus, we neglect this term, since it  cancels
when we subtract the zero field effective Lagrangian.

The remaining term in $I_1$ can also be written in terms of
$\xi(\kappa)$, as
\begin{eqnarray}
\int_{\frac{2ef}{\Lambda^2}}^\infty dy\,
\E^{-\frac{m^2}{ef} y} 
   \frac{\coth\left(y-\frac{ef}{\Lambda^2}\right)}{\sinh^2 y} &=&-\frac{1}{3}(1+\ln 2)
+2\kappa\,\xi(\kappa)\nonumber\\
&&+
\kappa^2\left[ \left(3-2\gamma+2\ln\left(\frac{\Lambda^2}{m^2}\right)\right)-4\frac{\Lambda^2}{m^2}\,\ln 2+ \left(4\ln 2-2\right)\,\frac{\Lambda^4}{m^4}\right]
\label{id2}
\end{eqnarray}
up to terms vanishing as the cutoff is removed. Note that the final term in (\ref{id2}) is proportional to $\kappa^2$, and so can be dropped as it leads to a field-independent contribution to
the effective Lagrangian.

So, putting everything together, we see that the entire two-loop
effective Lagrangian (\ref{decomp}) can be written in terms of the
function $\xi(\kappa)$:
\begin{eqnarray}
{\cal L}^{(2)}&=&\frac{\alpha^2}{(4\pi)^2} f^2\left\{
    2(A+B) \xi^\prime - 2(A+2B)\xi^2
-4\kappa\, \xi\left[(A+2B)\left(\ln\left(\frac{\Lambda^2}{m^2}\right)
-\gamma\right)+\left(\frac{3}{2}A+2B\right)\right]\right.\nonumber\\
&&\left. +\left[2(A+B)\left(\ln\left(\frac{\Lambda^2}{m^2}\right)
-\gamma\right) +\left(\frac{5}{6}A+B\right)-\left(\frac{5}{3} A+2 B\right)\,\ln 2\right]\right\}
   \label{2lgen}
\end{eqnarray}
The last term on the RHS of (\ref{2lgen}) is proportional to the bare Maxwell Lagrangian $f^2$, and so corresponds to the charge renormalization counterterm.
However, even after doing this charge renormalization there remains on the
RHS of (\ref{2lgen}) a logarithmic UV divergence  with a nontrivial field
dependence $\sim
\xi(\kappa)$. This term can be seen to contribute to mass
renormalization by noting that
\begin{equation}
-4 \kappa\, \xi =8\pi^2 \left[ 
  \left\{ \begin{array}{c} 1\\ -2 \end{array} \right\}
  \frac{m^2}{(ef)^2} \frac{\partial}{\partial m^2}{\cal
  L}^{(1)\,{\rm ren}}_{\text{\tiny $\left\{ \begin{array}{c}
  \text{sp}\\ \text{sc} \end{array} \right\}$}}
     - \frac{1}{8\pi^2} 
    \left\{ \begin{array}{c} 1\\ 0 \end{array} \right\} \right],
  \label{3.A.7}
\end{equation}
where we have used a combined notation for spinor (upper) and scalar
(lower) QED. The important difference arises from the last term
in Eq.~(\ref{3.A.7}): in the scalar case this term is zero, and the $\xi$
function is the only contribution required for a mass renormalization
of the one-loop Lagrangian,
\begin{equation}
{\cal L}^{(1)}(ef,m_{\text{R}}^2)= {\cal L}^{(1)}(ef,m^2)+\frac{\partial
  {\cal L}^{(1)}(ef,m^2)}{\partial m^2} \delta m^2, \label{3.A.8}
\end{equation}
where $m_{\text{R}}$ denotes the renormalized mass, and $\delta m^2$
is the mass renormalization counterterm. However, in the spinor case, the
$\xi$ function is not sufficient, but has to be supplemented by the last
term of Eq.~(\ref{3.A.7}) which accounts for the zero-mode contribution
in ${\cal L}^{(1)}_{\text{spinor}}$. 

Inserting the mass renormalization representation (\ref{3.A.7}) for $\kappa\,\xi$
into (\ref{2lgen}), we find that
the unrenormalized two-loop Lagrangian can finally be written as
\begin{eqnarray}
{\cal L}^{(2)}_{\text{\tiny $\left\{ \!\begin{array}{c} \text{sp}\\
       \text{sc} \end{array}\! \right\}$}}\!\!
&=&\!\!
  \frac{\alpha^2}{(4\pi)^2}\, f^2\, \left[ 2(A+B)\xi' -2(A+2B) \xi^2
       \right] \nonumber\\
&&\!\!+\frac{\alpha}{8\pi}
   \left\{ \begin{array}{c} 1\\ -2 \end{array} \right\}
   \left[ (A+2B)\left(\ln
\left(\frac{\Lambda^2}{m^2}\right)-\gamma\right) +\left(\frac{3A}{2}+2B
       \right)\right]
     m^2\frac{\partial}{\partial m^2} 
   {\cal L}^{(1)\,{\rm ren}}_{\text{\tiny $\left\{ \begin{array}{c}
\text{sp}\\
       \text{sc} \end{array} \right\}$}}\nonumber\\
&+&\!\!
  \frac{\alpha^2}{(4\pi)^2}\, f^2\, \left[   \left\{ \begin{array}{c} A\\ 2(A+B) \end{array} \right\}  \left(\ln\left(\frac{\Lambda^2}{m^2}\right)-\gamma\right) +
 \left\{ \begin{array}{c} -\frac{2A}{3}-B\\   \frac{5A}{6}+B\end{array} \right\}    -
\left(\frac{5A}{3}+2B\right)\ln 2
 \right] 
 \label{final}
\end{eqnarray}
This is our final result for the bare regularized two-loop Lagrangian,
written in a transparent way such that renormalization is almost
self-evident. The renormalized two-loop Lagrangian corresponds to the
first term in (\ref{final}); inserting $A=-4$ and $B=8$ in the spinor
case, and
$A=B=-1$ in the scalar case, we rediscover the results of
\cite{ds1,ds2,ds3} quoted in Eqs.~(\ref{2lsp},\ref{2lsc}). Note that the results of \cite{ds1,ds2,ds3} 
were derived using the world-line representation of the effective Lagrangian, so the result (\ref{final}) provides an independent confirmation.

The second
term in
(\ref{final}) represents the mass renormalization counter-term that has
to be added to the one-loop Lagrangian in the spirit of
Eq.~(\ref{3.A.8}). Here we can also read off the one-loop mass shift
(apart from ``tadpole'' contribution for the scalar case, as discussed
above),
\begin{equation}
\delta m^2_{\text{\tiny $\left\{ \begin{array}{c}
  \text{sp}\\ \text{sc} \end{array} \right\}$}}
  =\left\{ \begin{array}{c} 1 \\ 1/2 \end{array}
  \right\}\frac{3\alpha}{2\pi} \left[ \left(\ln
\left(\frac{\Lambda^2}{m^2}\right) -\gamma\right)+ 
    \left\{ \begin{array}{c} 5/6 \\ 7/6 \end{array} \right\} \right]\,m^2,
  \label{3.A.10}
\end{equation}
which agrees with independent one-loop computations using a
proper-time cutoff \cite{schwinger,tsai}.

The last term in (\ref{final}) correponds to the charge renormalization
counterterm which has to be added to the Maxwell Lagrangian in order to
renormalize the coupling and field strength (as in the one-loop case, we can
trade the UV cutoff scale $\Lambda$ for an arbitrary renormalization
scale $\mu$). Inserting the appropriate values for the coefficients $A$
and $B$ we find
\bear
\delta {\cal L}_{\text{\tiny $\left\{ \begin{array}{c} \text{sp}\\
       \text{sc} \end{array} \right\}$}}^{(2){\rm charge\,\, ren.}}=-
\frac{e^4}{64\pi^4}\, f^2 
  \left[\left(\ln\left(\frac{\Lambda^2}{m^2}\right)-\gamma\right)
     +  \left\{ \begin{array}{c} 4/3\\ 11/24 \end{array}
                      \right\} +\left\{ \begin{array}{c} 7/3\\ -11/12
\end{array} 
                      \right\}\ln 2 \right].
\label{chargeren}
\ear
As expected, from these charge renormalization terms we can read off the
correct two-loop $\beta$ function coefficients quoted in
Eqs.(\ref{spbeta},\ref{scbeta}).  

The two-loop origin of the mis-match (in the spinor case with a
self-dual background) between the
$\beta$ function coefficient and the strong-field behavior becomes
clear now: although the zero-mode contribution exerts no direct
influence on the IR behavior of the unrenormalized Lagrangian (cf.
Eq.~(\ref{2lsplagzm2})), the zero-mode contribution to the mass
renormalization term in Eq.~(\ref{3.A.7}) introduces another UV divergence
which, together with the overall UV divergence of the unrenormalized
Lagrangian, leads to the correct $\beta$ function. Whereas the overall
UV divergence contributes equally to the strong-field limit by the UV/IR
connection, the zero-mode UV divergence from the mass renormalization
does not affect the strong field limit. This is the subtle source of the
mis-match in the spinor case at two-loop. We stress again that this is
very different from the more familiar role of the zero-modes at
one-loop, as described in Section IIIB.

\section{Conclusions}
\label{conclusions}
\renewcommand{\theequation}{5.\arabic{equation}}
\setcounter{equation}{0}

We have analyzed the relation between the short-distance behavior and
the strong-field limit of QED with electromagnetic backgrounds. On the
one hand, the strong-field asymptotics of a renormalized QED effective
Lagrangian is generally determined by its infrared behavior. Since, on
the other hand, the terms which are relevant for the ultraviolet
behavior affect also the infrared simply for dimensional reasons,
quantum fluctuations induce an IR/UV interplay. In many instances,
this mechanism leads to an exact IR/UV correspondence between the
strong-field limit and the $\beta$ function. For instance in the case
of magnetic backgrounds or scalar QED, the strong-field limit can be
computed from the $\beta$ function and vice versa, as is also
suggested by an argument involving the trace anomaly. The necessary
condition for this exact IR/UV correspondence as well as the
trace-anomaly argument is the mass independence of the strong-field
limit or, phrased differently, the validity of the theory in the
massless limit.
 
In the case of spinor QED in a self-dual background, an apparent
discrepancy 
arises from a comparison of the strong-field behavior of the
effective Lagrangian with the behavior predicted by the
perturbative $\beta$ function and a naive application of the
trace-anomaly argument. The key to the resolution is the appearance of
zero modes for spinor QED in a self-dual background which invalidate a
direct massless limit. This is well understood at one-loop, but we
found that the role of the zero modes is rather different at two loop.
Indeed, one way to understand our two loop results is that there
is, in fact, only really a one-loop effect: at the two-loop level, the
zero modes do not introduce a new IR divergence, but enter instead
through the inevitable reappearance of the one-loop effective
Lagrangian via mass renormalization.
 
One motivation for our work is to prepare for future studies of
higher-loop calculations in QCD for quarks in a self-dual instanton
background. A great deal is known about this at one-loop
\cite{thooft,carlitz,leeleepac,kwon}, but not at the two loop level.
Many features of our QED analysis generalize to the instanton case
because it was primarily the self-duality of the background, rather
than its spacetime independence, which was most important. However,
one major difference is that in QED the internal photon propagator
does not feel the background field, while for the corresponding QCD
diagram the internal gluon propagator {\it does} couple to the
background (instanton) field. Here it would be interesting to make
connection with the QED and QCD analysis of the one-loop polarization
operator $\Pi_{\mu\nu}(Q^2)$ in a self-dual background, where the role
of the zero modes has also been studied \cite{smilga,finjord}.

Finally, we conclude with a discussion of a renormalization group (RG)
interpretation of our results, which gives another perspective to the
IR/UV connection in the renormalization of spinor and scalar QED in these
self-dual backgrounds. The discrepancy between the strong-field limit and
the $\beta$ function coefficients can be viewed from a different
perspective with the aid of a renormalization group (RG) equation for
the effective Lagrangian. The RG equation can be derived from the
statement that the renormalized Lagrangian is independent of the
renormalization scale $\mu$,
\begin{equation}
\mu\frac{d}{d\mu}\, {\cal L}(eF,\alpha,m;\mu)=0, \label{C.A}
\end{equation}
where all quantities are assumed to be renormalized. Equation
(\ref{C.A}) states that any shift in $\mu$ is compensated for by
corresponding shifts of the renormalized parameters. Since the product
$eF$ is RG invariant, it acts only as a spectator in the
following considerations and can be omitted from now on. Introducing
the anomalous mass dimension
\begin{equation}
\gamma_m=-\frac{\mu}{m} \frac{\partial m}{\partial \mu}, \label{C.B}
\end{equation}
the RG equation can be written as
\begin{equation}
\left( \mu\frac{\partial}{\partial\mu} + \beta^{(\alpha)}
    \frac{\partial}{\partial \alpha} -\gamma_m\,
    m\frac{\partial}{\partial m} \right)\, {\cal
    L}(\alpha,m;\mu)=0. \label{C.C}
\end{equation}
If the strong-field limit was mass-independent, we could drop the term
$\sim\gamma_m$ in Eq.~(\ref{C.C}) and read off the $\beta$ function
from this limit. In the self-dual spinor case, however, the
mass-dependence induced by the zero modes forces us to keep this term
even in the strong-field limit where ${\cal L}$ at one-loop is given
by
\begin{equation}
{\cal L}=-\frac{(ef)^2}{4\pi \alpha} -\frac{(ef)^2}{24\pi^2} \ln
\frac{ef}{m^2} +\frac{(ef)^2}{12\pi^2} \ln
\frac{m^2}{\mu^2}, \quad\text{for}\,\,\frac{ef}{m^2}\to \infty. \label{C.D}
\end{equation}
The first term is simply the renormalized Maxwell term. Inserting
Eq.~(\ref{C.D}) into Eq.~(\ref{C.C}) leads us to
\begin{equation}
\beta^{(\alpha)}_m=\frac{2}{3} \frac{\alpha^2}{\pi} +\gamma_m\,
\frac{\alpha^2}{\pi} +\dots, \label{C.E}
\end{equation}
where the dots represent higher-loop contributions. Here we appended
the subscript $m$ to the $\beta$ function in order to indicate the
mass dependence. Since $\gamma_m$ is of order $\alpha$, namely
\begin{equation}
\gamma_m=-\frac{\mu}{m}\frac{\partial m}{\partial \mu} =-\frac{1}{2}
\frac{\mu}{m^2} \frac{\partial \delta m^2}{\partial \mu} =-\frac{3}{2}
\frac{\alpha}{\pi} +\dots, \label{C.F}
\end{equation}
as can be read off from Eq.~(\ref{3.A.10}) by trading $\Lambda$ for
$\mu$, the mass dependence induces contributions to the $\beta_m$
function at the two-loop level and higher. Adding the standard
two-loop coefficient as obtained within a mass-independent scheme, we
find
\begin{equation}
\beta^{(\alpha)}_m=\frac{2}{3} \frac{\alpha^2}{\pi} + \frac{1}{2}
\frac{\alpha^3}{\pi^2} -\frac{3}{2} \frac{\alpha^3}{\pi^2} +{\cal O}(\alpha^4)
=\frac{2}{3} \frac{\alpha^2}{\pi} -\frac{\alpha^3}{\pi^2} + {\cal
  O}(\alpha^4), \label{C.G}
\end{equation}
so that the two-loop coefficient in the nomenclature used in
Sect.~\ref{strong} reads $\beta_{m,2}=-1/(32 \pi^4)$. This coefficient
matches perfectly with the two-loop strong-field limit of the
self-dual spinor case given in Eq.~(\ref{2lspinorsdasym}). 

We can interpret the coincidence in the following way: there is, in
fact, a correspondence between the strong-field limit and the $\beta$
function in the self-dual spinor case at two-loop; but this
correspondence applies only to the $\beta$ function $\beta_m$ of an
implicitly electron-mass-dependent regularization scheme. (Note that
the standard argument \cite{Gross:vu} of scheme-independence of the two-loop
coefficient holds for mass-independent schemes only.) This
mass-dependent scheme is natural in the self-dual spinor case because
of the presence of the zero modes which inhibit a direct massless
limit. 

This analysis can be performed at any loop order. In those cases where
the strong-field limit is mass-independent such as a magnetic
background, this analysis connects the $\beta$ function with the
strong-field limit coefficients and is well understood
\cite{ritus1,ritus2,ritus3,ditreu}. In the present case, however, such
an analysis connects the mass-dependent $\beta$ function, the
anomalous mass dimension and the strong-field limit with each
other. For instance, if the strong-field limit at $n$-loop order and
the anomalous dimension at ($n$-1)-loop order are known, we can
extract the $n$-loop mass-dependent $\beta_m$ function and also the
mass-independent $\beta$ function by virtue of the $n$-loop analogue
of Eq.~(\ref{C.E}). Aiming at an $n$-loop computation of the $\beta$
function, this is the same amount of information required as for a
magnetic background, but the computation for a self-dual background
will be much simpler.

As a further remark, let us point out that we have discussed possible
massless limits of QED always as continuous limits of massive theories
in this work. In this sense, a massless limit of the self-dual spinor
case does not exist because of the zero modes. This does not imply
that a massless formulation of the self-dual spinor case does not
exist at all. On the contrary, it is well possible that a massless
formulation exists but requires a different treatment similar to the
case of massless gluonic fluctuations in a self-dual Yang-Mills
background \cite{leutwyler,thooft}. For this, the integration over the
fermions has to be decomposed into zero-mode and non-zero-mode
fluctuations. The non-zero modes have to be integrated out first in a
background consisting of the constant self-dual field plus zero-mode
fluctuations. We expect that the non-zero-mode integration ``dresses''
the zero modes in such a way that they acquire a mass. Contrary to the
gluonic case, this mechanism requires a two-loop calculation, so that
an effective four-fermion coupling between the zero and nonzero modes
can be generated by photon exchange. In a self-dual background, the
non-zero modes will form a condensate which then gives a mass to
the zero modes because of this four-fermion interaction. A strong
evidence for this scenario is given by our observation that the zero
modes do not induce an IR divergence at the two-loop level. Once the
zero modes are lifted by this effective mass, they can finally be
integrated out. Since there is no further scale in this formulation,
the standard relation between the strong-field limit and the beta
function can be expected to hold for this theory once the zero modes
are integrated out. This explains also why the massless limit of the
massive case cannot be continuous because the strong-field limit
coefficient changes discontinuously in this limit.

\acknowledgments{We thank V. I. Ritus for discussions. GD acknowledges
  the US Department of Energy for support through grant
  DE-FG02-92ER40716.  GD and CS thank NSF and CONACyT for a US-Mexico
  collaborative research grant NSF-INT-0122615. HG thanks the
  University of Connecticut Research Foundation for support during a
  visit to UConn and acknowledges the Deutsche Forschungsgemeinschaft
  for support through grant Gi 328/1-1.}

\section{Appendix}
\label{appendix}
\renewcommand{\theequation}{A.\arabic{equation}}
\setcounter{equation}{0}

In this appendix we comment on the {\it finite} part of the mass
renormalization in (\ref{3.A.10}). This finite part is not relevant
for the main discussion of this paper, as it does not affect either the
$\beta$ function or the strong-field limit. However, it is crucial for
the derivation of the finite renormalized two-loop effective Lagrangians
(\ref{2lsp}) and (\ref{2lsc}) for spinor and scalar QED, respectively, in
a constant self-dual background. These two-loop results were first
derived in \cite{ds1,ds2} using the worldline formalism, and here we have
given an independent  derivation in Section IV of this current paper
using a conventional field theory diagrammatic approach. 

In order to fix the finite part of the mass renormalization mass shift,
one approach is to compare (\ref{3.A.10}) with an independent calculation
of the UV properties of the mass operator \cite{schwinger,tsai},
done in the same regularization scheme. For a constant magnetic
background, this scheme dependence of the finite mass shift has been
studied at two-loop in \cite{frss}. Another approach, implicit in
\cite{ritus1,ritus2,ritus3,lebedev}, is to demand that the leading growth rate of
the coefficients of the weak-field expansion of the two-loop renormalized
effective Lagrangian coincides (up to a factor of $\alpha\pi$) with the 
leading growth rate of the
coefficients of the weak-field expansion of the one-loop renormalized
effective Lagrangian. This ensures that the leading imaginary part of the
${\cal L}_{\rm eff}$, when the field is analytically continued to an
unstable regime, involves the same physical electron mass at two-loop as
at one-loop. This is because the leading imaginary parts go
like $\exp[-m^2\pi/(e |f|)]$, and these nonperturbative factors are
related to the leading divergence rate of the perturbative coefficients of
the (divergent) weak-field expansion in the standard way . Therefore, 
any mis-match between the leading growth rates at one-loop and
two-loop corresponds to a shifted value of $m^2$, and vice versa. This
gives an interesting ``nonperturbative'' definition of the renormalized
mass, which is completely compatible with the standard definition of the
mass through a renormalized perturbative Green's function
\cite{ritus1,ritus2,ritus3}. The correspondence of the leading growth
rates of the one-loop and two-loop weak-field expansions has been
confirmed numerically in
\cite{2leh} for the case of a constant electric background, and has been
confirmed analytically in \cite{ds3} for the case of a constant self-dual
background (with $f$ analytically continued $f\to if$).

From the calculation presented in Section IV, it is easy to see from
(\ref{3.A.8}) and (\ref{final}) that any finite shift in the finite parts
($5/6$ and $7/6$ for spinor and scalar, respectively) of the mass shift
in (\ref{3.A.10}) would introduce into the renormalized two-loop effective
Lagrangians on the first line of (\ref{final}) an additional term of the
form 
\bear
\delta {\cal L}^{(2)}
&\sim&\alpha\, m^2\frac{\partial}{\partial
m^2}{\cal L}^{(1)}\sim \alpha\,
\kappa\frac{\partial}{\partial\kappa}{\cal L}^{(1)}.
\label{finiteshift}
\ear
Now the one-loop effective Lagrangians in (\ref{1lspinorsd}) and
(\ref{1lscalarsd}) have divergent weak-field expansions of the form
\bear
{\cal L}^{(1)}=m^4 \sum_{n=2}^\infty \frac{c_n^{(1)}}{\kappa^{2n}},
\label{1lweak}
\ear
where the magnitude of the expansion coefficients grows factorially as
\cite{ds3}
\bear
|c_n^{(1)}|\sim \frac{\Gamma(2n-1)}{(2\pi)^{2n}}.
\label{1lgrowth}
\ear
Similarly, the two-loop effective Lagrangians in (\ref{2lsp}) and
(\ref{2lsc}) have divergent weak-field expansions of the form
\bear
{\cal L}^{(2)}=\alpha\pi\, m^4 \sum_{n=2}^\infty
\frac{c_n^{(2)}}{\kappa^{2n}},
\label{2lweak}
\ear 
where the magnitude of the expansion coefficients grows factorially as
\cite{ds3}
\bear
|c_n^{(2)}|\sim \frac{\Gamma(2n-1)}{(2\pi)^{2n}}.
\label{2lgrowth}
\ear
This leading growth rate is precisely the same as the one-loop rate in
(\ref{1lgrowth}), confirming Ritus's criterion at this order.

The two-loop results (\ref{2lsp}) and (\ref{2lsc}) for the on-shell
renormalized effective Lagrangians appear in the first line of
(\ref{final}), when the finite parts of the mass shifts are as
specified in (\ref{3.A.10}). If these finite parts were shifted, then
the renormalized effective Lagrangians would acquire a further shift
as in (\ref{finiteshift}). However, this extra piece clearly has the
wrong growth rate, with the magintude of the expansion coefficients
now growing like $\Gamma(2n)/(2\pi)^{2n}$, which is faster than the
one-loop growth rate in (\ref{1lgrowth}). Thus we see that we can
indeed uniquely implement Ritus's nonperturbative criterion as a means
to fix the physical renormalized mass, including the finite part of
the mass shift. And the result is completely consistent with the
finite parts found by standard perturbative means
\cite{schwinger,tsai}. We believe that the method described here is
not only of theoretical interest, but at higher loop orders might
actually be technically preferable to a direct calculation of the mass
shift.


\begin{thebibliography}{99}

\bibitem{ritus1}
V. I. Ritus, ``Lagrangian of an intense electromagnetic
field and quantum electrodynamics at short distances'', Zh. Eksp. Teor.
Fiz {\bf 69} (1975) 1517 [Sov. Phys. JETP {\bf 42} (1975) 774].

\bibitem{ritus2}
V. I. Ritus, ``Connection between strong-field quantum electrodynamics
with short-distance quantum electrodynamics'',
Zh. Eksp. Teor. Fiz {\bf 73} (1977) 807
[Sov. Phys. JETP {\bf 46} (1977) 423].

\bibitem{ritus3}
V. I. Ritus, ``The Lagrangian Function of an Intense Electromagnetic
Field'', in {\it Proc. Lebedev Phys. Inst.} Vol. {\bf 168}, {\it Issues
in Intense-field Quantum Electrodynamics}, V. I. Ginzburg, ed., (Nova
Science Pub., NY 1987); ``Effective Lagrange function of intense
electromagnetic field in QED,'' in Proceedings of the conference 
{\it Frontier Tests of QED and Physics of the Vacuum}, 
E.Zavattini et al (Eds), (Heron Press, Sofia,
1998), [arXiv:hep-th/9812124].

\bibitem{crewther}
R.~J.~Crewther,
``Nonperturbative Evaluation Of The Anomalies In Low-Energy Theorems,''
Phys.\ Rev.\ Lett.\  {\bf 28}, 1421 (1972);
M.~S.~Chanowitz and J.~R.~Ellis,
``Canonical Anomalies And Broken Scale Invariance,''
Phys.\ Lett.\ B {\bf 40}, 397 (1972).



\bibitem{anomaly}
S.~L.~Adler, J.~C.~Collins and A.~Duncan,
``Energy-Momentum-Tensor Trace Anomaly In Spin 1/2 Quantum
Electrodynamics,'' Phys.\ Rev.\ D {\bf 15}, 1712 (1977);
J.~C.~Collins, A.~Duncan and S.~D.~Joglekar,
``Trace And Dilatation Anomalies In Gauge Theories,''
Phys.\ Rev.\ D {\bf 16}, 438 (1977).

\bibitem{nielsen}
N.~K.~Nielsen,
``The Energy Momentum Tensor In A Nonabelian Quark Gluon Theory,''
Nucl.\ Phys.\ B {\bf 120}, 212 (1977).


\bibitem{matinian}
I.~A.~Batalin, S.~G.~Matinian and G.~K.~Savvidy,
``Vacuum Polarization By A Source - Free Gauge Field,''
Sov.\ J.\ Nucl.\ Phys.\  {\bf 26}, 214 (1977)
[Yad.\ Fiz.\  {\bf 26}, 407 (1977)];
S.~G.~Matinian and G.~K.~Savvidy,
``Vacuum Polarization Induced By The Intense Gauge Field,''
Nucl.\ Phys.\ B {\bf 134}, 539 (1978).

\bibitem{pagels} H.~Pagels and E.~Tomboulis,
``Vacuum Of The Quantum Yang-Mills Theory And Magnetostatics,''
Nucl.\ Phys.\ B {\bf 143}, 485 (1978).

\bibitem{leutwyler}
H.~Leutwyler,
``Vacuum Fluctuations Surrounding Soft Gluon Fields,''
Phys.\ Lett.\ B {\bf 96}, 154 (1980).
``Constant Gauge Fields And Their Quantum Fluctuations,''
Nucl.\ Phys.\ B {\bf 179}, 129 (1981).

\bibitem{minkowski}
P.~Minkowski,
``On The Ground State Expectation Value Of The Field Strength Bilinear
In Gauge Theories And Constant Classical Fields,'' Nucl.\ Phys.\ B {\bf
177}, 203 (1981).

\bibitem{elizalde}
E.~Elizalde,
``Effective Lagrangian For Ordinary Quarks In A Background Field,''
Nucl.\ Phys.\ B {\bf 243}, 398 (1984);
E.~Elizalde and J.~Soto,
``Zeta Regularized Lagrangians For Massive Quarks In Constant Background Mean Fields,''
Annals Phys.\  {\bf 162}, 192 (1985).


\bibitem{fujikawa} K.~Fujikawa,
``A Nondiagramatic Calculation Of One Loop Beta Function In QCD,''
Phys.\ Rev.\ D {\bf 48}, 3922 (1993).

\bibitem{hansson} J.~Grundberg and T.~H.~Hansson,
``The QCD trace anomaly as a vacuum effect (The vacuum is a medium is
the message!),'' Annals Phys.\  {\bf 242}, 413 (1995)
[arXiv:hep-th/9407139].

\bibitem{giesbook}
W.~Dittrich and H.~Gies,
{\it Probing the quantum vacuum. Perturbative effective action approach in
quantum electrodynamics and its application},  Springer Tracts Mod.\
Phys.\  {\bf 166}  (2000).

\bibitem{shifman}
M.~A.~Shifman and A.~I.~Vainshtein,
``Operator Product Expansion And Calculation Of The Two Loop
Gell-Mann-Low Function,'' Sov.\ J.\ Nucl.\ Phys.\  {\bf 44}, 321 (1986)
[Yad.\ Fiz.\  {\bf 44}, 498 (1986)].

\bibitem{ss}
M.~G.~Schmidt and C.~Schubert,
``Multiloop calculations in the string inspired formalism: The Single spinor loop in QED,''
Phys.\ Rev.\ D {\bf 53}, 2150 (1996)
[arXiv:hep-th/9410100].



\bibitem{schwarz}
A.~S.~Schwarz,
``On Regular Solutions Of Euclidean Yang-Mills Equations,''
Phys.\ Lett.\ B {\bf 67}, 172 (1977);
R.~Jackiw and C.~Rebbi,
``Spinor Analysis Of Yang-Mills Theory,''
Phys.\ Rev.\ D {\bf 16}, 1052 (1977).

\bibitem{ds1}
G.~V.~Dunne and C.~Schubert,
``Closed-form two-loop Euler-Heisenberg Lagrangian in a self-dual 
background,'' Phys.\ Lett.\ B {\bf 526}, 55 (2002)
[arXiv:hep-th/0111134].

\bibitem{ds2}
G.~V.~Dunne and C.~Schubert,
``Two-loop self-dual Euler-Heisenberg Lagrangians. I: Real part and 
helicity amplitudes,'' JHEP {\bf 0208}, 053 (2002)
[arXiv:hep-th/0205004].

\bibitem{ds3}
G.~V.~Dunne and C.~Schubert,
``Two-loop self-dual Euler-Heisenberg Lagrangians. II: Imaginary part 
and Borel analysis,'' JHEP {\bf 0206}, 042 (2002)
[arXiv:hep-th/0205005].

\bibitem{broadhurst} 
D.~J.~Broadhurst,
``Dimensionally continued multi-loop gauge theory,''
arXiv:hep-th/9909185.

\bibitem{eh}
W. Heisenberg and H. Euler, ``Folgerungen aus der
Diracschen Theorie des Positrons'', Z. Phys. {\bf 98} (1936) 714.

\bibitem{schwinger}
J. Schwinger, ``On gauge invariance and vacuum polarization'', Phys. Rev.
{\bf 82} (1951) 664.

\bibitem{ditreu}
W.~Dittrich and M.~Reuter,
{\it Effective Lagrangians In Quantum Electrodynamics},
Springer Lect.\ Notes Phys.\  {\bf 220}  (1985).

\bibitem{thooft}
G.~'t Hooft,
``Computation Of The Quantum Effects Due To A Four-Dimensional 
Pseudoparticle,'' Phys.\ Rev.\ D {\bf 14}, 3432 (1976)
[Erratum-ibid.\ D {\bf 18}, 2199 (1978)].


\bibitem{dadda}
A.~D'Adda and P.~Di Vecchia,
``Supersymmetry And Instantons,''
Phys.\ Lett.\ B {\bf 73}, 162 (1978).

\bibitem{duff}
M.~J.~Duff and C.~J.~Isham,
``Selfduality, Helicity, And Coherent States In Nonabelian Gauge
Theories,'' Nucl.\ Phys.\ B {\bf 162}, 271 (1980);
``Selfduality, Helicity, And Supersymmetry: The Scattering Of Light By
Light,'' Phys.\ Lett.\ B {\bf 86}, 157 (1979).

\bibitem{mahlon} G. Mahlon, ``One loop multiphoton helicity amplitudes'', 
Phys. Rev. D {\bf 49} (1994) 2197, hep-ph/9311213;
%
Z.~Bern, L.~J.~Dixon, M.~Perelstein and J.~S.~Rozowsky,
``Multi-leg one-loop gravity amplitudes from gauge theory,''
Nucl.\ Phys.\ B {\bf 546}, 423 (1999)
[arXiv:hep-th/9811140];
%
Z. Bern, A. De Freitas,
L. Dixon, A. Ghinculov and H. L. Wong, ``QCD and QED corrections to
light-by-light scattering'', JHEP {\bf 0111} (2001) 031, hep-ph/0109079; 
%
Z.~Bern, A.~De Freitas and L.~Dixon,
``Two-loop helicity amplitudes for gluon gluon scattering in QCD and  
supersymmetric Yang-Mills theory,''
JHEP {\bf 0203}, 018 (2002)
[arXiv:hep-ph/0201161];
T.~Binoth, E.~W.~Glover, P.~Marquard and J.~J.~van der Bij,
``Two-loop corrections to light-by-light scattering in supersymmetric  
QED,''
JHEP {\bf 0205}, 060 (2002)
[arXiv:hep-ph/0202266].

\bibitem{bardeen} W. Bardeen, ``Self-dual Yang-Mills theory, integrability
and multiparton amplitudes'', Prog. Theor. Phys. Suppl. {\bf 123} (1996)
1; D. Cangemi, ``Self-duality and maximally helicity
violating QCD amplitudes'', Int. J. Mod. Phys. {\bf A12}
(1997) 1215, [arXiv:hep-th/9610021].

\bibitem{abramowitz} M. Abramowitz and I. Stegun, {\it Handbook of
Mathematical Functions}, (Dover, 1998).

\bibitem{ck}
L.~S.~Brown, R.~D.~Carlitz and C.~Lee,
``Massless Excitations In Instanton Fields,''
Phys.\ Rev.\ D {\bf 16}, 417 (1977);
R.~D.~Carlitz and C.~Lee,
``Physical Processes In Pseudoparticle Fields: The Role Of Fermionic
Zero Modes,'' Phys.\ Rev.\ D {\bf 17}, 3238 (1978).

\bibitem{carlitz}
R.~D.~Carlitz and D.~B.~Creamer,
``Light Quarks And Instantons,''
Annals Phys.\  {\bf 118}, 429 (1979).


\bibitem{leeleepac}
C.~Lee, H.~W.~Lee and P.~Y.~Pac,
``Calculation Of One Loop Instanton Determinants Using Propagators With
Space-Time Dependent Mass,'' Nucl.\ Phys.\ B {\bf 201}, 429 (1982).

\bibitem{tsai}
W.~Y.~Tsai,
``Magnetic Bremsstrahlung And Modified Propagation Function. Spin-0 
Charged Particles In A Homogeneous Magnetic Field,''
Phys.\ Rev.\ D {\bf 8}, 3460 (1973).



\bibitem{kwon} O.~K.~Kwon, C.~Lee and H.~Min,
``Massive field contributions to the QCD vacuum tunneling amplitude,''
Phys.\ Rev.\ D {\bf 62}, 114022 (2000)
[arXiv:hep-ph/0008028].

\bibitem{smilga}
M.~S.~Dubovikov and A.~V.~Smilga,
``Analytical Properties Of The Quark Polarization Operator In An External
Selfdual Field,'' Nucl.\ Phys.\ B {\bf 185} (1981) 109.

\bibitem{finjord} J.~ Finjord, ``Quarks in a constant self-dual Euclidean
background field (I)'', Nucl. Phys. {\bf B 194} (1982) 77; 
``Quarks in a constant self-dual Euclidean
background field (II)'', Nucl. Phys. {\bf B 222} (1983) 507.
 
\bibitem{Gross:vu}
G.~M.~Avdeeva, A.~A.~Belavin and A.~P.~Protogenov,
``On Possibility Of Existence Of A Finite Charge In The Quantum Field Theory,''
Yad.\ Fiz.\  {\bf 18}, 1309 (1973);
D.~J.~Gross,
``Applications Of The Renormalization Group To High-Energy Physics,''
{\it  In *Les Houches 1975, Proceedings, Methods In Field Theory*, Amsterdam 1976, 141-250}.

\bibitem{frss}
M.~Reuter, M.~G.~Schmidt and C.~Schubert,
``Constant external fields in gauge theory and the spin 0, 1/2, 1 path  integrals,''
Annals Phys.\  {\bf 259}, 313 (1997)
[arXiv:hep-th/9610191];
D.~ Fliegner, M.~ Reuter, M.~ G.~ Schmidt and C. ~Schubert, 
``The two-loop Euler-Heisenberg Lagrangian in dimensional
renormalization'',  Teor. Mat. Fiz. {\bf 113} (1997) 289
[Theor. Math. Phys. {\bf 113} (1997) 1442].

\bibitem{lebedev}
S.~L.~ Lebedev and  V~.I.~ Ritus, ``Virial representation of the imaginary part of the Lagrange function of the electromagnetic field'', Sov. Phys. JETP {\bf 59} (1984) 237 [Zh. Eksp. Teor. Fiz. {\bf 86} (1984) 408].

\bibitem{2leh}
G.~V.~Dunne and C.~Schubert,
``Two-loop Euler-Heisenberg QED pair-production rate,''
Nucl.\ Phys.\ B {\bf 564}, 591 (2000)
[arXiv:hep-th/9907190].

 
\end{thebibliography}
\end{document}